\documentclass [11pt,a4paper] {article}

\usepackage[cp1252]{inputenc}

\usepackage{amssymb}
\usepackage{amsmath}
\usepackage{amsfonts,amssymb}
\usepackage[dvips]{graphicx}
\usepackage{bbm}
\usepackage{enumerate}

\usepackage{amsthm}

\usepackage{cancel}

\DeclareMathAlphabet{\mathpzc}{OT1}{pzc}{m}{it}

\def\SmallColSep{\setlength{\arraycolsep}{1pt}}

\begin{document}

\title{The problem of dispersion-free probabilities in Gleason-type theorems for a two-dimensional Hilbert space}

\author{Arkady Bolotin\footnote{$Email: arkadyv@bgu.ac.il$\vspace{5pt}} \\ \textit{Ben-Gurion University of the Negev, Beersheba (Israel)}}

\maketitle

\begin{abstract}\noindent As it is known, Gleason’s theorem is not applicable for a two-dimensional Hilbert space since in this situation Gleason’s axioms are not strong enough to imply Born’s rule thus leaving room for a dispersion-free probability measure i.e., one that has only values 0 and 1. To strengthen Gleason’s axioms one must add at least one more assumption. But, as it is argued in the present paper, alternatively one can give up the lattice condition lying in the foundation of Gleason’s theorem. Particularly, the lattice structure based on the closed linear subspaces in the Hilbert space could be weakened by the requirement for the meet operation to exist only for the subspaces belonging to commutable projection operators. The paper demonstrates that this weakening can resolve the problem of the dispersion-free probability measure in the case of a qubit.
\\

\end{abstract}

\section{Introduction and preliminaries}  

\noindent Recall that the set of two or more \textit{nontrivial} (that is, different from $\hat{0}$ and $\hat{1}$) projection operators $\hat{P}_A$, $\hat{P}_B$, … on a Hilbert space $\mathcal{H}$ associated with a quantum system is called \textit{a context $\Sigma$} if the following requirements are satisfied:\smallskip

\begin{equation}  
   \hat{P}_A
   ,
   \hat{P}_B
   \in
   \Sigma
   \;\;
   \implies
   \;\;
   \left\{
      \begin{array}{l}
         \hat{P}_A
         \wedge
         \hat{P}_B
         =
         \hat{P}_A
         \hat{P}_B
         =
         \hat{P}_B
         \hat{P}_A
         =
         \hat{0}
         \\ 
         \hat{P}_A
         \vee
         \hat{P}_B
         =
         \hat{P}_A
         +
         \hat{P}_B
      \end{array}
   \right.
   \;\;\;\;  ,
\end{equation}

\begin{equation}  
   \sum_{\hat{P} \in \Sigma}
   \hat{P} 
   =
   \hat{1}
   \;\;\;\;  ,
\end{equation}
\smallskip

\noindent where the logical operations $\hat{P}_A \wedge \hat{P}_B$ and $\hat{P}_A \vee \hat{P}_B$ correspond, respectively, to the intersection and linear span of the ranges of the projection operators, i.e., $\mathrm{ran}(\hat{P}_A)$ and $\mathrm{ran}(\hat{P}_B)$.\\

\noindent Let $P$ stand for the experimental proposition corresponding to the projection operator $\hat{P} \!\in\! \Sigma$ and let $\mathrm{Pr}[P(|\Psi\rangle)] \in [0,1]$ be the probability of $P$ assuming the value of true in the quantum state of the system $|\Psi\rangle$ such that\smallskip

\begin{equation} \label{GL1} 
   \mathrm{Pr}
   \left[
      P_A(|\Psi\rangle)
      \vee
      P_B(|\Psi\rangle)
   \right]
   =
   \mathrm{Pr}\left[P_A(|\Psi\rangle)\right]
   +
   \mathrm{Pr}\left[P_B(|\Psi\rangle)\right]
   \;\;\;\;  ,
\end{equation}

\begin{equation} \label{GL2} 
   \mathrm{Pr}
   \left[
      I(|\Psi\rangle)
   \right]
   =
   1
   \;\;\;\;  ,
\end{equation}
\smallskip

\noindent\noindent where $\vee$ denotes the operation of logical disjunction on the values of the propositions $P_A$ and $P_B$ in the state $|\Psi\rangle$, and $I$ represents the trivially-true proposition corresponding to $\mathrm{ran}(\hat{1})$. As long as these axioms hold, Gleason's theorem states that there is a unique density operator $\rho$ on the Hilbert space $\mathcal{H}$ with $\dim(\mathcal{H}) \ge 3$ such that\smallskip

\begin{equation}  
   \mathrm{Pr}
   \left[
      P(|\Psi\rangle)
   \right]
   =
   \mathrm{Tr}
   \!\!
   \left[
      \rho
      \hat{P}
   \right]
   \;\;\;\;  .
\end{equation}
\smallskip

\noindent Providing $\hat{P} =|\Psi\rangle\langle\Psi|$ and the system is prepared in the pure state $\rho = |\Phi\rangle\langle\Phi|$, it gives Born's rule:\smallskip

\begin{equation}  
   \mathrm{Pr}
   \left[
      P(|\Psi\rangle)
   \right]
   =
   \left|
      \langle\Phi
      |
      \Psi\rangle
   \right|^2
   \;\;\;\;  .
\end{equation}
\smallskip

\noindent However, Gleason’s theorem is not applicable for a two-dimensional Hilbert space.\\

\noindent To see this, consider the sublattice $\mathcal{M}(\Sigma)$ of the Hilbert lattice $\mathcal{L}(\mathbb{C}^2)$, i.e., the orthomodular lattice formed by the closed linear subspaces of the two-dimensional Hilbert space $\mathcal{H}=\mathbb{C}^2$. The sublattice $\mathcal{M}(\Sigma)$ is a subset of $\mathcal{L}(\mathbb{C}^2)$, that is a lattice with the same meet $\wedge$ and join $\vee$ operations as $\mathcal{L}(\mathbb{C}^2)$, namely,\smallskip

\begin{equation}  
   \mathcal{M}(\Sigma)
   =
   \Bigg\{
      \{0\}
      ,
      \left\{
         \!\left[
            \begingroup\SmallColSep
            \begin{array}{r}
               a \\
               a
            \end{array}
            \endgroup
         \right]\!
      \right\}
      ,
      \left\{
         \!\left[
            \begingroup\SmallColSep
            \begin{array}{r}
               a \\
              -a
            \end{array}
            \endgroup
         \right]\!
      \right\}
      ,
      \left\{
         \!\left[
            \begingroup\SmallColSep
            \begin{array}{r}
               ia \\
               a
            \end{array}
            \endgroup
         \right]\!
      \right\}
      ,
      \left\{
         \!\left[
            \begingroup\SmallColSep
            \begin{array}{r}
               a \\
              ia
            \end{array}
            \endgroup
         \right]\!
      \right\}
      ,
      \left\{
         \!\left[
            \begingroup\SmallColSep
            \begin{array}{r}
               a \\
               0
            \end{array}
            \endgroup
         \right]\!
      \right\}
      ,
      \left\{
         \!\left[
            \begingroup\SmallColSep
            \begin{array}{r}
               0 \\
               a
            \end{array}
            \endgroup
         \right]\!
      \right\}
      ,
      \mathbb{C}^2
   \Bigg\}
   \;\;\;\;  ,
\end{equation}
\smallskip

\noindent where $a\in \mathbb{R}$. The elements of this sublattice are the ranges of the projection operators $\hat{P}^{(Q)}_n$ on $\mathbb{C}^2$, i.e.,\smallskip

\begin{equation}  
   \left\{
      \!\left[
         \begingroup\SmallColSep
         \begin{array}{r}
            \cdot \\
            \cdot
         \end{array}
         \endgroup
      \right]\!
   \right\}
   =
   \mathrm{ran}(\hat{P}^{(Q)}_n)
   \;\;\;\;  ,
\end{equation}
\smallskip

\noindent that are defined by the formula\smallskip

\begin{equation}  
   \hat{P}^{(Q)}_n
   =
   \frac{1}{2}
      \left[
         \begin{array}{l l}
            1+(-1)^n (\delta_{Q0}-\delta_{Q3})
            &
            (-1)^n (-\delta_{Q1}+i\delta_{Q2})
            \\
            (-1)^n (-\delta_{Q1}-i\delta_{Q2})
            &
            1+(-1)^n (\delta_{Q0}+\delta_{Q3})
         \end{array}
      \right]
   \;\;\;\;  ,
\end{equation}
\smallskip

\noindent in which $\delta_{ab}$ is the Kronecker delta, $n \in \{1,2\}$ and $Q \in \{0,1,2,3\}$. According to this formula,\smallskip

\begin{equation}  
   \{0\}
   =
   \mathrm{ran}(\hat{P}^{(0)}_1)
   =
   \mathrm{ran}(\hat{0})
   \;\;\;\;  ,
\end{equation}

\begin{equation}  
   \mathbb{C}^2
   =
   \mathrm{ran}(\hat{P}^{(0)}_2)
   =
   \mathrm{ran}(\hat{1})
   \;\;\;\;  ,
\end{equation}
\smallskip

\noindent where $\mathrm{ran}(\hat{0})$ and $\mathrm{ran}(\hat{1})$ are the trivial closed subspaces.\\

\noindent Let us define the elementary event $E^{(Q)}_n$ for $Q \neq 0$ as the following outcome of the experiment:\smallskip

\begin{equation}  
   E^{(Q)}_n
   \equiv
   \left\{
      |\Psi\rangle:\;|\Psi\rangle \in \mathrm{ran}(\hat{P}^{(Q)}_n)
   \right\}
   \;\;\;\;  .
\end{equation}
\smallskip

\noindent This event occurs if the vector $|\Psi\rangle$, associated with the pure state of \textit{a qubit} (a two-state quantum system such as an one-half spin particle), is found in the nontrivial closed subspace $\mathrm{ran}(\hat{P}^{(Q)}_n)$ and, as a result, the proposition corresponding to the nontrivial projection operator $\hat{P}^{(Q)}_n$ (e.g., ``spin along the $Q$-axis is up'') is verified (i.e., has the value of the truth).\\

\noindent Since all the vectors associated with physically meaningful states of the qubit must differ from zero, no $|\Psi\rangle$ resides in $\mathrm{ran}(\hat{0})$. Thus, it must be\smallskip

\begin{equation}  
   \left\{
      |\Psi\rangle:\;|\Psi\rangle \in \mathrm{ran}(\hat{0})
   \right\}
   =
   \varnothing
   \;\;\;\;  .
\end{equation}
\smallskip

\noindent Also, given that all $|\Psi\rangle \in \mathcal{H}$, it must be\smallskip

\begin{equation}  
   \left\{
      |\Psi\rangle:\;|\Psi\rangle \in \mathrm{ran}(\hat{1})
   \right\}
   =
   \Omega
   \;\;\;\;  ,
\end{equation}
\smallskip

\noindent where $\Omega$ denotes the set of all possible outcomes of the experiment.\\

\noindent Suppose that the qubit is prepared in the pure state $|\Psi^{(Q)}_n\rangle$ lying in the range of the nontrivial projection operator $\mathrm{ran}(\hat{P}^{(Q)}_n)$. Then, in the experiment on the measurement of $\hat{P}^{(Q)}_n$, the set $\Omega$ contains only the event $E^{(Q)}_n$:\smallskip

\begin{equation}  
   E^{(Q)}_n
   =
   \left\{
      |\Psi^{(Q)}_n\rangle:\;|\Psi^{(Q)}_n\rangle \in \mathrm{ran}(\hat{P}^{(Q)}_n)
   \right\}
   =
   \Omega
   \;\;\;\;  .
\end{equation}
\smallskip

\noindent Now, consider the event $E^{(R)}_m$ different from $E_n^{(Q)}$ (which means that either $R \neq Q$ or $m \neq n$ or both) consisting of the outcome\smallskip

\begin{equation}  
   E^{(R)}_m
   =
   \left\{
      |\Psi^{(Q)}_n\rangle
      :
      \;
      |\Psi^{(Q)}_n\rangle
      \in
     \mathrm{ran}(\hat{P}^{(Q)}_n)
     \wedge
     \mathrm{ran}(\hat{P}^{(R)}_m)
   \right\}
   \;\;\;\;  ,
\end{equation}
\smallskip

\noindent where the subspace $\mathrm{ran}(\hat{P}^{(Q)}_n) \wedge \mathrm{ran}(\hat{P}^{(R)}_m)$ (if it exists) is the meet of $\mathrm{ran}(\hat{P}^{(Q)}_n)$ and $\mathrm{ran}(\hat{P}^{(R)}_m)$. This event can occur if the proposition corresponding to the projection operator $\hat{P}^{(R)}_m$ (e.g., ``spin along the $R$-axis is up'') is verified in the state $|\Psi^{(Q)}_n\rangle$. Next, recall that in the sublattice $\mathcal{M}(\Sigma)$ the meet exists for every pair $\mathrm{ran}(\hat{P}^{(Q)}_n)$ and $\mathrm{ran}(\hat{P}^{(R)}_m)$, namely,\smallskip

\begin{equation}  
  \mathrm{ran}(\hat{P}^{(Q)}_n)
  \wedge
  \mathrm{ran}(\hat{P}^{(R)}_m)
   =
   \mathrm{ran}(\hat{P}^{(Q)}_n)
  \cap
  \mathrm{ran}(\hat{P}^{(R)}_m)
   =
   \{0\}
  \;\;\;\;  ,
\end{equation}
\smallskip

\noindent where $\cap$ is the set-theoretical intersection. For all $R \neq Q$ this entails\smallskip

\begin{equation}  
   E^{(R)}_m
   =
   \left\{
      |\Psi^{(Q)}_n\rangle
      :
      \;
      |\Psi^{(Q)}_n\rangle
      \in
      \{0\}
   \right\}
   =
   \varnothing
   \;\;\;\;  ,
\end{equation}

\begin{equation}  
   E^{(R)}_1
   \cap
   E^{(R)}_2
   =
   \left\{
      |\Psi^{(Q)}_n\rangle
      :
      \;
      |\Psi^{(Q)}_n\rangle
      \in
      \mathrm{ran}(\hat{P}^{(Q)}_n)
      \wedge
      \left(
         \mathrm{ran}(\hat{P}^{(R)}_1)
         \cap
         \mathrm{ran}(\hat{P}^{(R)}_2)
      \right)
   \right\}
   =
   \varnothing
   \;\;\;\;  ,
\end{equation}

\begin{equation}  
   E^{(R)}_1
   \cup
   E^{(R)}_2
   =
   \left\{
      |\Psi^{(Q)}_n\rangle
      :
      \;
      |\Psi^{(Q)}_n\rangle
      \in
      \mathrm{ran}(\hat{P}^{(Q)}_n)
      \wedge
      \left(
         \mathrm{ran}(\hat{P}^{(R)}_1)
         +
         \mathrm{ran}(\hat{P}^{(R)}_2)
      \right)
   \right\}
   =
   \Omega
   \;\;\;\;  .
\end{equation}
\smallskip

\noindent Providing $\Pr[E^{(Q)}_n]$ is the probability of the event $E^{(Q)}_n$ in a manner that $\Pr[\varnothing]=0$ and $\Pr[\Omega]=1$, one gets the following \textit{dispersion-free} (i.e., having only values 0 and 1) probability measure:\smallskip

\begin{equation} \label{DF1} 
   \Pr[ E^{(R)}_1 \cup E^{(R)}_2 ]
   =
   \Pr[ E^{(R)}_1 ]
   +
   \Pr[ E^{(R)}_2 ]
   =
   1
   \;\;\;\;  ,
\end{equation}

\begin{equation} \label{DF2} 
   \Pr[ E^{(R)}_{m \in \{1,2\}} ]
   =
   0
   \;\;\;\;  .
\end{equation}
\smallskip

\noindent \noindent The problem with this measure, however, is that it allows $\Pr[ E^{(R)}_{m} ]$ -- i.e., the probability of finding the state $|\Psi_m^{(R)}\rangle \in \mathrm{ran}(\hat{P}^{(R)}_m)$ when the system has been prepared in the state $|\Psi_n^{(Q)}\rangle \in \mathrm{ran}(\hat{P}^{(Q)}_n)$ -- to be equal to zero which contradicts Born's rule stating that $\Pr[ E^{(R)}_{m} ] = |\langle\Psi_n^{(Q)}|\Psi_m^{(R)}\rangle|^2$ must differ from zero.\\

\noindent To overcome this problem, one may replace the projection operators $\hat{P}^{(Q)}_n$ by \textit{effects $\hat{E}_i$}, that is, self-adjoint operators bounded between $\hat{0}$ and $\hat{1}$, namely,\smallskip

\begin{equation}  
   \hat{0} \le \hat{E}_i \le \hat{1}
   \;\;
   \implies
   \;\;
   0 \le \langle\Psi|\hat{E}_i|\Psi\rangle \le 1
   \;\;\;\;   ,
\end{equation}
\smallskip

\noindent and demand\smallskip

\begin{equation}  
   \Pr[\hat{E}_1 + \hat{E}_2 + \dots]
   =
   \Pr[\hat{E}_1]
   +
   \Pr[\hat{E}_2]
   +
   \dots
   \;\;\;\;    
\end{equation}
\smallskip

\noindent in case of\smallskip

\begin{equation}  
   \hat{E}_1 + \hat{E}_2 + \dots
   \le
   1
   \;\;\;\;   ,
\end{equation}
\smallskip

\noindent where the commutativity is not necessary to hold \cite{Busch03}. Then, the fact that the orthogonality constraint $\hat{E}_i \hat{E}_{j \neq i} = \hat{E}_{j \neq i} \hat{E}_i = 0$ is no longer required will preclude dispersion-free probabilities of the events relating to measurements of the properties associated with $\hat{E}_i$.\\

\noindent Yet, the replacement of the qubit projection operators $\hat{P}^{(Q)}_n$ by the effects $\hat{E}_i$ (corresponding to elements of \textit{a positive-operator-valued measure}, POVM, \cite{Busch97}) brings about a new problem, namely, the problem of the interpretation of the effects $\hat{E}_i$. To be sure, unlike the projection operators $\hat{P}^{(Q)}_n$ whose number is limited (by the dimension of the Hilbert space associated with the system), the number of the effects $\hat{E}_i$ is unlimited. More importantly, one always obtains the same result when performing two consecutive verifications of the proposition corresponding to the projection operator $\hat{P}^{(Q)}_n$, while this need not be true for the properties associated with the effects $\hat{E}_i$ \cite{Nielsen}. As they are not repeatable, the effects can be regarded as imperfect observations.\\

\noindent Another way to avoid the equality (\ref{DF2}) is to show that it is \textit{illogical}, i.e., not justifiable from rationality principles \cite{Benavoli}.\\

\noindent However, this approach is not conceptually neutral. That is, the pertinence of the rationality principles to quantum mechanics strongly depends on the interpretation of the state vector $|\Psi\rangle$. Thus, in the Bayesian approach to quantum mechanics \cite{Caves, Morgan, Savage, Finetti, Bernardo}, probabilities – and therefore the state vector $|\Psi\rangle$ – represent an agent's degrees of belief (which can be rational or not rational), rather than objective properties of physical systems (as it is assumed in accordance with an ontic interpretation of $|\Psi\rangle$). In view of that, the idea of excluding the dispersion-free probability measure from consideration based on the rationality principles can be rationalized only within QBism, i.e., the Bayesian interpretation of quantum mechanics.\\

\noindent Then again, to avoid the expression (\ref{DF2}), one can give up the lattice condition. Concretely, the Hilbert lattice structure could be weakened by the requirement for the meet operation $\wedge$ to exist only for the ranges belonging to the commutable projection operators. The present paper demonstrates that such a weakening can resolve the problem of the dispersion-free probability measure in the case of a two-dimensional Hilbert space.\\

\section{The structure of invariant-subspace lattices}  

\noindent The said weakening of the Hilbert lattice structure can be obtained by introducing the collection of invariant-subspace lattices $\mathcal{L}(\Sigma^{(1)})$, $\mathcal{L}(\Sigma^{(2)})$ and $\mathcal{L}(\Sigma^{(3)})$ in place of the Hilbert sublattice $\mathcal{M}(\Sigma)$.\\

\noindent To that end, consider the set $\mathcal{L}(\Sigma^{(Q)})$ that includes only such subspaces $\mathcal{H}^\prime$ from $\mathcal{M}(\Sigma)$ that are invariant under each nontrivial projection operator $\hat{P}^{(Q)}_n$ from the context $\Sigma^{(Q)} = \{ \hat{P}^{(Q)}_1, \hat{P}^{(Q)}_2 \}$. This means that the image of every vector $|\Psi\rangle$ in $\mathcal{H}^\prime$ under $\hat{P}^{(Q)}_n \in \Sigma^{(Q)}$ remains within $\mathcal{H}^\prime$, i.e.,\smallskip

\begin{equation}  
   \hat{P}^{(Q)}_n \mathcal{H}^\prime
   =
   \left\{
      |\Psi\rangle \in \mathcal{H}^\prime
      :\;
      \hat{P}^{(Q)}_n |\Psi\rangle
   \right\}
   \subseteq
   \mathcal{H}^\prime
   \;\;\;\;  .
\end{equation}
\smallskip

\noindent The elements of the set $\mathcal{L}(\Sigma^{(Q)})$, explicitly,\smallskip

\begin{equation}  
   \mathcal{L}(\Sigma^{(Q)})
   =
   \left\{
      \{0\}
      ,\,
      \mathrm{ran}(\hat{P}^{(Q)}_1)
      ,\,
      \mathrm{ran}(\hat{P}^{(Q)}_2)
      ,\,
      \mathbb{C}^2
   \right\}
   \;\;\;\;  ,
\end{equation}
\smallskip

\noindent form \textit{the invariant-subspace lattice}, a complete complemented distributive lattice (a Boolean algebra) \cite{Davey, Radjavi}.\\ 

\noindent As it is obvious, each $\mathcal{L}(\Sigma^{(Q)})$ only contains the closed linear subspaces belonging to the mutually commutable projection operators. Hence, the nontrivial $\mathrm{ran}(\hat{P}^{(Q)}_n)$ can only belong to the invariant-subspace lattice $\mathcal{L}(\Sigma^{(Q)})$, while the nontrivial $\mathrm{ran}(\hat{P}^{(R)}_m)$ – only to $\mathcal{L}(\Sigma^{(R)})$.\\ 

\noindent One can observe here that if the lattices $\mathcal{L}(\Sigma^{(1)})$, $\mathcal{L}(\Sigma^{(2)})$ and $\mathcal{L}(\Sigma^{(3)})$ are pasted together at the trivial subspaces $\{0\}$ and $\mathbb{C}^2$, then the resulted logic will be the Hilbert sublattice $\mathcal{M}(\Sigma)$. Thus, the Hilbert sublattice $\mathcal{M}(\Sigma)$ can be thought of as \textit{the pasting together of the invariant-subspace lattices from the collection} $\{\mathcal{L}(\Sigma^{(Q)})\}_Q$.\\

\noindent  Without such pasting – i.e., giving up the assumption of the Hilbert lattice – the closed linear subspaces belonging to the incommutable projection operators cannot meet each other within the structure of $\{\mathcal{L}(\Sigma^{(Q)})\}_Q$. In symbols,\smallskip

\begin{equation}  
   \mathrm{ran}(\hat{P}_n^{(Q)})
   \in
   \mathcal{L}(\Sigma^{(Q)})
   ,\,
   \mathrm{ran}(\hat{P}_m^{(R)})
   \in
   \mathcal{L}(\Sigma^{(R)})
   \,
   \implies
   \,
   \mathrm{ran}(\hat{P}_n^{(Q)})
   \;\cancel{\;\wedge\;}\;
   \mathrm{ran}(\hat{P}_m^{(R)})
   \;\;\;\;  ,
\end{equation}
\smallskip

\noindent where the cancelation of the meet operation $\wedge$ indicates that it cannot be defined for such elements (recall that the meet is defined as an operation on pairs of elements from one lattice \cite{Davey}).\\

\noindent Since the sets\smallskip

\begin{equation}  
   E^{(R)}_m
   =
   \left\{
      |\Psi^{(Q)}_n\rangle
      :\;
      |\Psi^{(Q)}_n\rangle
      \in
      \mathrm{ran}(\hat{P}_n^{(Q)})
      \;\cancel{\;\wedge\;}\;
      \mathrm{ran}(\hat{P}_m^{(R)})
   \right\}
   \;\;\;\;   
\end{equation}
\smallskip

\noindent cannot be defined, the events $E^{(R)}_m$ must be \textit{objectively indeterminate}, despite the fact that the intersection and union of these events can be determined, namely, $E^{(R)}_1 \!\cap E^{(R)}_2 = \varnothing$ and $E^{(R)}_1 \!\cup E^{(R)}_2 = \Omega$.\\

\noindent This implies that within the structure of $\{\mathcal{L}(\Sigma^{(Q)})\}_Q$, the closed linear subspaces of the two-dimensional Hilbert space do not admit probabilities having only values 0 and 1.\\

\section{Concluding remarks}  

\noindent When one has no other information than that the mutually exclusive events $E^{(R)}_1$ and $E^{(R)}_2$ are expected if the system is prepared in state $|\Psi^{(Q)}_n\rangle$, one is justified in assigning each the equal probability, that is,\smallskip

\begin{equation}  
   \Pr[ E^{(R)}_m ]
   =
   \Pr
   \left[
   \left\{
      |\Psi^{(Q)}_n\rangle
      :\;
      |\Psi^{(Q)}_n\rangle
      \in
      \mathrm{ran}(\hat{P}_n^{(Q)})
      \;\cancel{\;\wedge\;}\;
      \mathrm{ran}(\hat{P}_m^{(R)})
   \right\}
   \right]
   =
   \frac{1}{2}
   \;\;\;\;  .
\end{equation}
\smallskip

\noindent This implies Born’s rule for the qubit\smallskip

\begin{equation} \label{J} 
   \Pr[ E^{(R)}_{m} ]
   =
   \left|
   \langle\Psi^{(Q)}_n
   |
   \Psi^{(R)}_m\rangle
   \right|^2
   =
   \frac{1}{2}
   \;\;\;\;  .
\end{equation}
\smallskip

\noindent It was noticed in the Introduction that Gleason's theorem \cite{Gleason}  is not applicable for the two-dimensional Hilbert space $\mathbb{C}^2$ because in the said case Gleason's axioms (\ref{GL1}) and (\ref{GL2})  are not strong enough to imply Born's rule thus leaving room for the equality (\ref{DF2}). So, to strengthen Gleason's axioms one must add at least one more assumption \cite{Hall}.\\

\noindent But, as it has been shown in the present paper, alternatively, one can weaken the structure of the Hilbert lattice $\mathcal{L}(\mathbb{C}^2)$ by introducing the invariant-subspace lattices $\mathcal{L}(\Sigma^{(1)})$, $\mathcal{L}(\Sigma^{(2)})$ and $\mathcal{L}(\Sigma^{(3)})$ in place of the sublattice $\mathcal{M}(\Sigma)$ of $\mathcal{L}(\mathbb{C}^2)$. As a consequence of this weakening, no assumption additional to Gleason’s axioms is necessary to imply Born’s rule (\ref{J}).\\

\section*{Acknowledgements}

\noindent The author wishes to thank the anonymous referee for the helpful feedback and insights.\\

\bibliographystyle{References}

\end{document}